\documentclass{article}
\usepackage{spconf,amsmath,graphicx,amssymb,makecell,multirow}
\title{ADDITIONAL SHARED DECODER ON SIAMESE MULTI-VIEW ENCODERS\\FOR LEARNING ACOUSTIC WORD EMBEDDINGS}
\name{Myunghun Jung, Hyungjun Lim, Jahyun Goo, Youngmoon Jung, and Hoirin Kim}
\address{School of Electrical Engineering, KAIST, Daejeon, Republic of Korea}
\begin{document}
%\ninept
%
\maketitle
% --------------------------------------------------------------------------
%           Abstract
% --------------------------------------------------------------------------
\begin{abstract}
Acoustic word embeddings --- fixed-dimensional vector representations of arbitrary-length words --- have attracted increasing interest in query-by-example spoken term detection.
Recently, on the fact that the orthography of text labels partly reflects the phonetic similarity between the words' pronunciation, a multi-view approach has been introduced that jointly learns acoustic and text embeddings.
It showed that it is possible to learn discriminative embeddings by designing the objective which takes text labels as well as word segments.
In this paper, we propose a network architecture that expands the multi-view approach by combining the Siamese multi-view encoders with a shared decoder network to maximize the effect of the relationship between acoustic and text embeddings in embedding space.
Discriminatively trained with multi-view triplet loss and decoding loss, our proposed approach achieves better performance on acoustic word discrimination task with the WSJ dataset, resulting in 11.1\% relative improvement in average precision.
We also present experimental results on cross-view word discrimination and word level speech recognition tasks.
\end{abstract}
% --------------------------------------------------------------------------
%           Intex Terms
% --------------------------------------------------------------------------
% --------------------------------------------------------------------------
%           1. Introduction
% --------------------------------------------------------------------------
\section{INTRODUCTION}
\label{sec:intro}

Query-by-example spoken term detection (QbE-STD) is the task of retrieving a spoken query from a set of speech utterances.
With the increasing use of smart devices that can interact through the user's voice (e.g. Amazon Echo, Google Home, Apple Siri), the QbE-STD has drawn interest as a technique that can be applied to wake-up or command word detection, search engine, etc.

In earlier works, approaches to compare a spoken query with speech utterances directly were proposed.
Feature vectors were calculated from the spoken query or speech utterances at frame-level, which were merged into a feature matrix.
The dynamic time warping (DTW) was used to measure the similarity between the feature matrices \cite{sakoe1990dynamic, hazen2009query, zhang2009unsupervised}.
Even though DTW is a very intuitive method, its matrix operation leads to high computational cost for each retrieval process.
In addition, there are a lot of potential target speech utterances.

As alternatives to approaches based on DTW, approaches to represent a word as a single vector, so-called acoustic word embedding \cite{levin2013fixed, chen2015query}, have been introduced.
Acoustic word embeddings are fixed-dimensional vector representations of arbitrary-length words, which are differentiated from semantic word embeddings \cite{mikolov2013distributed, chung2018speech2vec} in the way that they reflects phonetic information.
Once two words are represented as acoustic word embeddings, it is very easy to measure the similarity between them through a simple vector operation.

Most deep learning approaches using a word as input unit essentially involve the out-of-vocabulary (OOV) problem.
In order to handle OOV words in the practical application of acoustic word embeddings, several studies training a Siamese network with a triplet loss \cite{bengio2014word, kamper2016deep, settle2016discriminative, settle2017query, yuan2018learning} have been conducted to learn phonetic similarity between word segments.
From the weak supervision that indicates if two word segments are of the same class or not, the network can map words which have similar phonetic properties onto close distributions in embedding space while mapping ones which have different phonetic properties onto distant distributions.
When the QbE-STD adopts acoustic word embeddings, it shows better performance than the DTW-based.

In case of given a transcribed speech dataset, a supervised learning method using the weak supervision does not make the best use of the label information.
In \cite{he2016multi}, W. He et al. focused on the fact that the orthography of text labels naturally reflects the phonetic similarity between the words' pronunciation and proposed a multi-view approach for jointly learning acoustic and text embeddings where word segments and text labels were used as two different input views of Siamese encoder networks.
This approach modified the weak supervision and Siamese network to suit a multi-view setting.

In this paper, we propose an advanced network architecture that expands \cite{he2016multi} by combining a decoder network to the Siamese multi-view encoders.
This additional decoder is shared and coupled with the acoustic and text encoders individually and composes an encoder-decoder and an autoencoder structure.
In the autoencoder where input is a text label and output is the reconstructed text label, the text embeddings are induced to have more representative capability in embedding space.
On the other hand, the encoder-decoder is trained to be able to decode the original text label from the acoustic embedding.
It makes the acoustic embeddings to learn underlying phonetic information, resulting in normalizing speech variances such as gender, age, tone, and intonation.
Also, when aligning distributions of acoustic and text embeddings discriminatively in the common space, the shared decoder can support this alignment.
Experimental results demonstrate that our proposed approach can learn more discriminative acoustic and text embeddings than previous work on acoustic word discrimination and cross-view word discrimination tasks.

% --------------------------------------------------------------------------
%           2. Multi-view Approach
% --------------------------------------------------------------------------
\section{MULTI-VIEW APPROACH}
\label{sec:multi}

First of all, we need to clarify some terms.
The single-view means that only a word segment is used as input.
Also, we distinguish the acoustic word embeddings of the multi-view approach into two subsets, acoustic and text embeddings, in accordance with their input data type.

In this section, we analyze how acoustic embeddings jointly learned with text embeddings in the multi-view approach \cite{he2016multi} can show better performance than ones in the general single-view approach.
This analysis has not been done properly in other works.

\subsection{Single-view approach}
\label{ssec:subsingle}

\begin{figure}[t]
\begin{minipage}[b]{1.0\linewidth}
  \centering
  \centerline{\includegraphics[width=31mm]{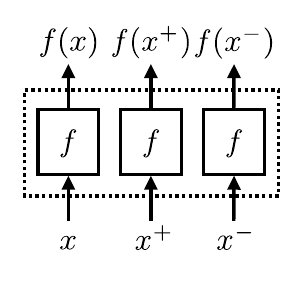}}
  \centerline{(a) Single-view approach}%\medskip
\end{minipage}
\begin{minipage}[b]{1.0\linewidth}
  \centering
  \centerline{\includegraphics[width=64mm]{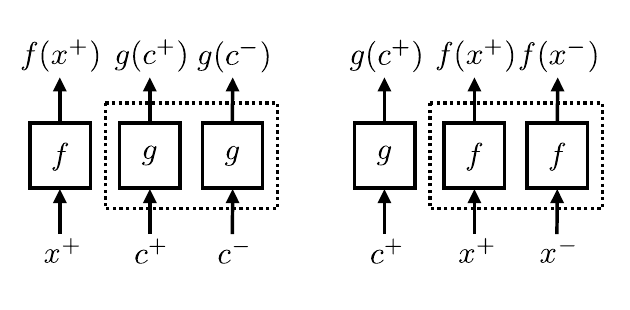}}
  \centerline{(b) Multi-view approach}%\medskip
\end{minipage}
\caption{Examples of triplet and Siamese encoder network for (a) single-view approach and (b) multi-view approach.}
\label{fig:siamese}
\end{figure}

In the single-view approach, a Siamese encoder network is trained with the weak supervision indicating that ``word segments $x$ and $x^+$ are samples of the same word'' and ``$x$ and $x^-$ are samples of different words''.
In Fig.\ref{fig:siamese}.(a), a triplet $(x, x^+, x^-)$ enters the encoder $f$, where the dotted line means the Siamese network, and acoustic word embeddings $f(x)$, $f(x^+)$, and $f(x^-)$ are extracted.
The main objective makes the distance between $f(x)$ and $f(x^+)$ smaller than the distance between $f(x)$ and $f(x^-)$ by a margin, which is optimized by the following triplet loss:
\begin{equation}
\Bigl[ m + d \left( f(x), f(x^+) \right) - d \left( f(x), f(x^-) \right) \Bigr]_+,
\end{equation}
where $\left [ a \right ]_+$ denotes the rectifier function $max(a, 0)$, the distance function $d(\cdot, \cdot)$ measures the distance between the two embeddings, and $m$ is the margin.
The training process of adjusting relative distances between embeddings makes the embeddings of the same word close to each other so that they can be distributed in one cluster in the embedding space.

\subsection{Multi-view approach}
\label{ssec:submulti}

In the multi-view approach, a text label is used as another kind of input view.
Like the single-view approach, weak supervision and Siamese network are used, but they should be modified to suit the multi-view setting.
When pairs of a word segment and a text label $(x, c)$ are given, the multi-view weak supervision indicates that ``$x^+$ and $c^+$ are samples of the same word'', ``$x^+$ and $c^-$ are samples of different words'', and ``$c^+$ and $x^-$ are samples of different words''.
Noticing that $x^-$ and $c^-$ are not chosen from one pair, three pairs $(x^+, c^+)$, $(x^-, \sim)$, and $(\sim, c^-)$ are used at a time.
Here $\sim$ denotes the unused sample.
As shown in Fig.\ref{fig:siamese}.(b), two triplets $(x^+, c^+, c^-)$ and $(c^+, x^+, x^-)$ that can be taken from this multi-view weak supervision enter the acoustic encoder $f$ and text encoder $g$.
In both cases, the Siamese network is applied to samples having the same view, not to all three samples of the triplet.
According to these multi-view triplet cases, the main objective consists of two parts ($obj^0 + obj^2$ in \cite{he2016multi}).
One makes the distance between $f(x^+)$ and $g(c^+)$ smaller than the distance between $f(x^+)$ and $g(c^-)$ by a margin, and the other makes the distance between $g(c^+)$ and $f(x^+)$ smaller than the distance between $g(c^+)$ and $f(x^-)$ by the margin.
This objective is optimized by the following two multi-view triplet losses:
\begin{equation} \label{eq:multi1}
\Bigl[ m + d \left( f(x^+), g(c^+) \right) - d \left( f(x^+), g(c^-) \right) \Bigr]_+,
\end{equation}
\begin{equation} \label{eq:multi2}
\Bigl[ m + d \left( g(c^+), f(x^+) \right) - d \left( g(c^+), f(x^-) \right) \Bigr]_+.
\end{equation}

As an anchor component of a triplet always has a different view from their positive and negative components, distances are indirectly adjusted through cross alignment between the acoustic and text embeddings.
Although the direct distance adjustment like the single-view approach is not achieved, embeddings can be effectively clustered in embedding space due to the uniqueness of text labels.
The encoder $f$ does not output the same acoustic embeddings even if input word segments express the same word, because speech appears as different instances every time.
In contrast, the text label of the word is unique and it means that the encoder $g$ outputs only one text embedding for one word.
At every training step, this unique text embeddings act as pivot points for the acoustic embeddings to be easily centralized, which is expressed in Eq.\ref{eq:multi2}.

The ultimate goal of learning acoustic word embeddings is to find inherent phonetic information for each word and to group similar ones in embedding space.
From this point of view, the single-view approach gathers embeddings of each word into individual clusters by extracting the common characteristics which, however, should be founded in the process of looking at many and various word samples without knowing where to focus on.
On the other hand, in the multi-view approach, it is assumed that each text embedding already represents inherent phonetic information.
Therefore, what networks need to do is simply to make acoustic embeddings be close to each text embedding.
At the same time, text embeddings are also trained to work as better reference points, so that the initial assumption becomes more influential.
Thus, the multi-view approach can learn more discriminative embeddings by utilizing text labels more effectively than the weak supervision of the single-view approach.
We use this multi-view approach as a baseline for our proposed approach.

% --------------------------------------------------------------------------
%           3. Proposed Approach
% --------------------------------------------------------------------------
\section{PROPOSED APPROACH}
\label{sec:proposed}

Contrary to single-view supervised learning based on a Siamese network, unsupervised learning methods for acoustic word embeddings mainly use an autoencoder structure \cite{chung2016audio, audhkhasi2017end, wang2018segmental, kamper2019truly}.
The architecture consists of an encoder network that extracts acoustic word embeddings from word segments and a decoder network that reconstructs the input segments from the embeddings.
Since one output word segment must be generated from only one embedding, embeddings are trained to compress the most essential information of given inputs.

An encoder-decoder structure is a general framework, which was used for learning semantic word embeddings \cite{chung2018speech2vec} or machine translation \cite{cho2014learning, sutskever2014sequence}.
It is the same as the autoencoder in that it consists of an encoder network and a decoder network, but it is distinguished in that it generates a different view of output from the input data.
Although input and output data are obtained from different views, they refer to the fundamentally equivalent content.
Therefore, the embeddings should be trained to represent underlying information that is shared between input and output data.

The most important part of the multi-view approach we analyzed was the assumption that the text embeddings represent inherent phonetic information of the words.
Although text embeddings move into better reference points at every training step, this is done by the incidental effect of the multi-view triplet loss, especially Eq.\ref{eq:multi1}.
In order to further improve the performance, it was necessary to introduce an additional objective to consolidate the assumption.
Thus, we paid attention to the role of the autoencoder in learning acoustic word embeddings and thought that it could be used for the enhancement of the text embeddings.
We also found that combining the decoder used in the autoencoder with the acoustic encoder makes it possible to take advantage of the encoder-decoder structure.
Based on these ideas, we have motivated a study that fully utilizes the autoencoder and encoder-decoder structure together to assist in learning acoustic and text embeddings.
In a similar study \cite{zhu2018siamese}, Z. Zhu et al. showed that acoustic word embeddings can be learned by combining the Siamese network of the single-view approach with the autoencoder structure.

\subsection{Network architecture}
\label{ssec:subnetwork}

\begin{figure}[t]
\begin{minipage}[b]{1.0\linewidth}
  \centering
  \centerline{\includegraphics[width=\linewidth]{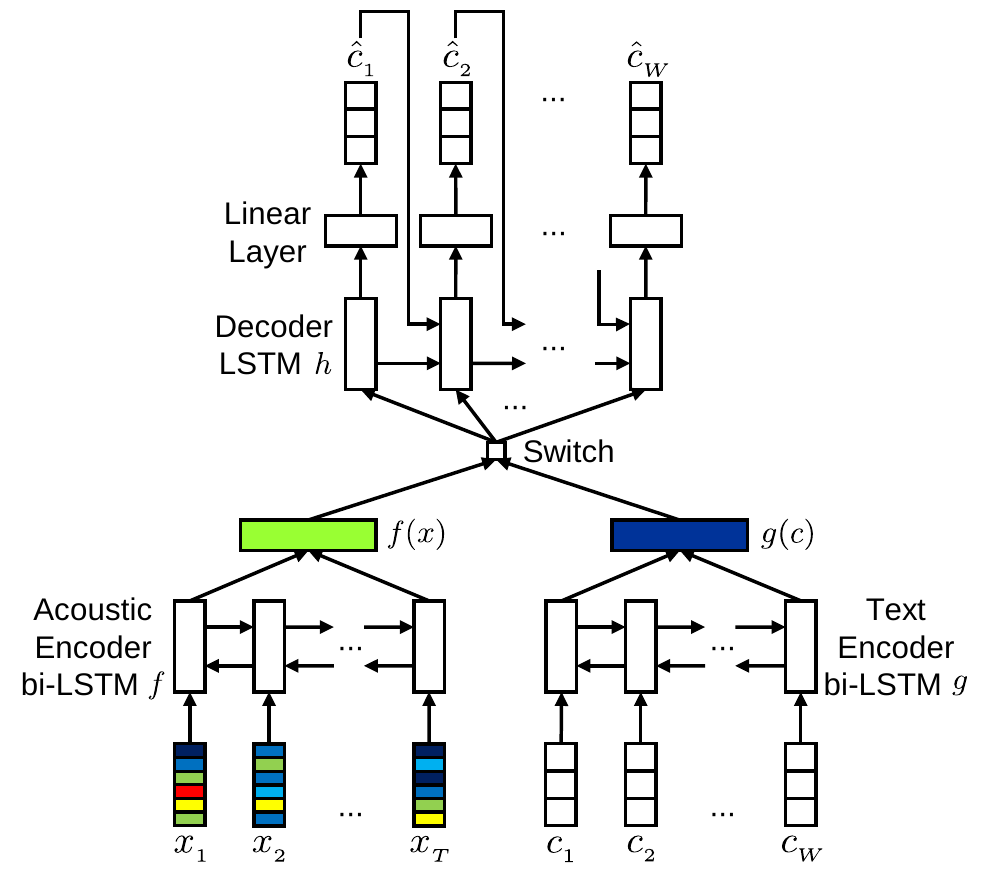}}
\end{minipage}
\caption{Illustration of proposed network architecture.}
\label{fig:architecture}
\end{figure}

In this paper, we propose an advanced network architecture that expands the multi-view approach in Sec.\ref{ssec:submulti} by combining a decoder network to the Siamese multi-view encoders.
As shown in Fig.\ref{fig:architecture}, this additional decoder $h$ is coupled and shared with the acoustic encoder $f$ and text encoder $g$ individually, so that composes acoustic-to-text encoder-decoder $(f \leftrightarrow h)$ and text-to-text autoencoder $(g \leftrightarrow h)$.
Given a pair of a word segment and a text label $(x, c)$, the encoder $f$ takes $x = \left \{ x_t \right \}^{T}_{t = 1}$ and outputs the acoustic embedding $f(x)$, where $x_t$ is the acoustic feature vector at frame $t$ and $T$ is the length of the $x$.
Also, the encoder $g$ takes $c = \left \{ c_w \right \}^{W}_{w = 1}$ and outputs the text embedding $g(c)$, where $c_w$ is the $w$-th character one-hot vector and $W$ is the length of the $c$.
Through the switch, either $f(x)$ or $g(c)$ is fed into the decoder $h$ generating a predicted or reconstructed text label $\hat{c} = \left \{ \hat{c}_y \right \}^{W}_{y = 1}$, where $\hat{c}_y$ is the probability distribution vector over all character classes for the $y$-th character. 

Since acoustic feature vectors and character vectors express sequential data, we implement the encoder $f$ and $g$ with multi-layer bidirectional long short term memory (bi-LSTM) networks \cite{hochreiter1997long, schuster1997bidirectional}.
At the output layer, the last hidden states of two directions are concatenated to form an embedding.
For the decoder $h$, a multi-layer unidirectional LSTM is used, but previous output $\hat{c}_{y-1}$ is used as an auxiliary input to calculate $\hat{c}_y$ at each step $y = 2, 3, ..., W$.
The hidden states of the output layer are transformed into lower dimension vectors through the fully-connected linear layer, which has the same number of output nodes as the character classes. Then $\hat{c}$ is calculated by the softmax operation.

\subsection{Training objective}
\label{ssec:subobjective}

In training of the proposed network architecture, a new objective is used in addition to the objective used in the multi-view approach.
The new objective corresponding to the decoder $h$ is that a decoded output $\hat{c}$ should be identical to text label $c$ regardless of whether the input embedding is $f(x)$ or $g(c)$.
This objective is optimized by the following decoding loss which is the sum of two cross-entropy losses: 
\begin{multline}
L_{decoding} = \sum_{i=1}^{N} \biggl( - \sum_{y=1}^{W} \Bigl( c_y^{i,+} \cdot log \left( \hat{c}_y \mid h(f(x^{i,+})) \right) +\\
c_y^{i,+} \cdot log \left( \hat{c}_{y} \mid h(g(c^{i,+})) \right) \Bigr) \biggr),
\end{multline}
where $(x^{i,+}, c^{i,+})$ is the $i$-th paired input data, $N$ is the size of training mini-batch, and $\cdot$ is element-wise dot product.

By the original role of the autoencoder, we can let the text embedding itself learn the identity of the word.
Deeply related to the uniqueness of text labels, the decoding loss maximizes the representative capability of the text embeddings in embedding space.
Also, by letting the shared decoder generate the same output from the paired acoustic and text embedding, these embeddings are tightly aligned.
Moreover, one target output is predicted from word segments having the same text label, allowing the acoustic embeddings to learn inherent phonetic information between two input views and to normalize speech variances.
This normalization effect is the result of the alignment and learning underlying information, although the word segments exist in various instances.

The overall training loss $L_{total}$ is the sum of multi-view triplet loss and decoding loss as follows:
\begin{equation}
L_{total} = L_{triplet} + \alpha L_{decoding},
\end{equation}
where $\alpha$ is a hyper-parameter which weights the decoding loss $L_{decoding}$, and $L_{triplet}$ is the sum of multi-view triplet losses from Eq.\ref{eq:multi1}, \ref{eq:multi2}.
$L_{triplet}$ is as follows:
\begin{multline}
L_{triplet} =\\
\sum_{i=1}^{N} \biggl( \Bigl[ m + d \left( f(x^{i,+}), g(c^{i,+}) \right) -
d \left( f(x^{i,+}), g(c^{i,-}) \right) \Bigr]_+ +\\
\Bigl[ m + d \left( g(c^{i,+}), f(x^{i,+}) \right) - d \left( g(c^{i,+}), f(x^{i,-}) \right) \Bigr]_+ \biggr),
\end{multline}
where $(x^{i,-}, \sim)$ and $(\sim, c^{i,-})$ are uniformly sampled negative input pairs from all of the differently labeled pairs in the training set according to the multi-view weak supervision.
In this paper we use the cosine distance, $d(\vec{p}, \vec{q}) = 1 - \frac{\vec{p} \cdot \vec{q}}{\left \| \vec{p} \right \| \left \| \vec{q} \right \|}$. 

% --------------------------------------------------------------------------
%           4. Experiments
% --------------------------------------------------------------------------
\section{EXPERIMENTS AND RESULTS}
\label{sec:experiments}

\subsection{Evaluation tasks}
\label{ssec:subtask}

Our original purpose is to improve the performance of the QbE-STD task, but it can be substituted with a word discrimination task on the condition that the word boundary information is known.
To measure the performance, we consider next two word discrimination tasks which are applicable in the multi-view setting.

The first task is \emph{acoustic word discrimination}, where we are given two word segments to determine whether they match or not.
This task is equivalent to the objective of the single-view approach and has been used in prior papers \cite{kamper2016deep, settle2016discriminative, settle2017query, yuan2018learning, chung2016audio, kamper2019truly}.
We regard this task as our main evaluation task for training the proposed and baseline network architectures.

The second task is \emph{cross-view word discrimination} corresponding to an audio-text QbE-STD where we retrieve a text query from a set of speech utterances or a spoken query from a set of text documents.
If a word segment and text label are given, we have to determine whether they indicate the same word or not.
This is very useful in that two different types of data can be easily integrated into one process.

In these two tasks, the cosine distance between embeddings of given two word expressions (word segment or text label) is calculated.
If the distance is below a threshold, we decide they are same, otherwise they are different words.
For the performance measure, we use the average precision (AP) which is the area under the precision-recall curve generated by sweeping the threshold.

\subsection{Dataset}
\label{ssec:subdataset}

The data used for experiments was drawn from the Continuous Speech Recognition Wall Street Journal Corpus (WSJ) \cite{paul1992design} Phase \MakeUppercase{\romannumeral 1} and Phase \MakeUppercase{\romannumeral 2}, specifically si\_tr\_s for the training set, si\_dt\_05 for the development set, and si\_et\_05, si\_et\_h2 for the test set.
All utterances were segmented into $(x, c)$ pairs by the forced alignment of the transcriptions from the GMM-HMM speech recognizer trained on the same training set using the open-source Kaldi toolkit \cite{povey2011kaldi}.
The number of pairs in training, development, and test set are 629897, 8616, and 9194 where the number of unique words are 13365, 1867, and 1728.
Having 42 million possible combinations of all pairs in the test set, there are 341932 matched cases and 41918289 unmatched cases.

We represented the acoustic feature vector $x_t$ using 40-dimensional Mel-filterbank energy which was calculated from 25 ms frame with 15 ms overlap.
The 26-dimensional character one-hot vector $c_w$ was generated by text normalizing that converts capitals and numbers into lower cases and removes quotation and punctuation, e.g. ``It's 7 o'clock.'' $\rightarrow$ ``its seven oclock'' $\rightarrow$ \{i/t/s\}\{s/e/v/e/n\}\{o/c/l/o/c/k\}.

\begin{table}[t]
\caption{Effect of various $\alpha$ in the development set AP for initial model and tuned model.}
\begin{center}
\begin{tabular}{c|cccccc}
\Xhline{3\arrayrulewidth}
$\alpha$ & 0 & 0.01 & 0.05 & 0.1 & 0.3 & 0.6 \\
\Xhline{3\arrayrulewidth}
Initial & 0.747 & 0.765 & 0.780 & \textbf{0.787} & 0.778 & 0.768 \\
\hline
Tuned & & 0.814 & 0.817 & \textbf{0.830} & 0.825 & 0.817 \\
\Xhline{3\arrayrulewidth}
\end{tabular}
\end{center}
\label{Tab:alphatable}
\end{table}

\begin{figure}[t]
\begin{minipage}[b]{1.0\linewidth}
  \centering
  \centerline{\includegraphics[width=\linewidth]{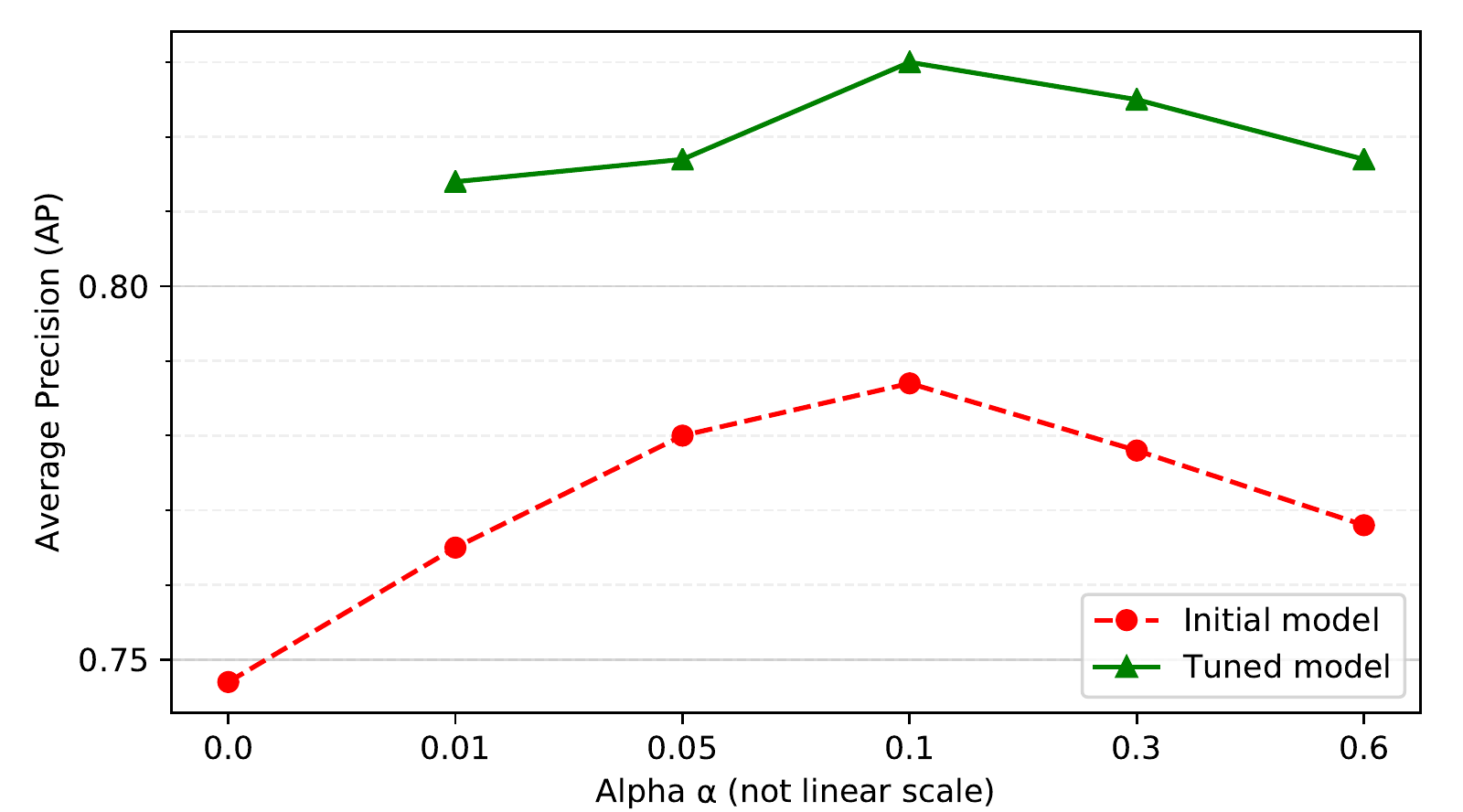}}
\end{minipage}
\caption{Tendency of varying the hyper-parameter $\alpha$ for initial model and tuned model.}
\label{fig:alphavalue}
\end{figure}

\subsection{Experimental setup}
\label{ssec:subsetup}

Before the experiment, we implemented the baseline multi-view approach \cite{he2016multi} and trained it with the model and dataset provided by the authors to verify the performance improvement compared to the single-view approaches \cite{kamper2016deep, settle2016discriminative}.
Then we established our initial model parameters as the same with the retuned baseline model on the WSJ dataset.

As the initial model for our proposed approach, 2-layer bi-LSTMs with 512 hidden units per direction were used for acoustic and text encoders.
Their weights were randomly initialized.
For the additional shared decoder which does not exist in the baseline architecture, we used a 2-layer LSTM with 512 hidden units.
The last states of each forward directional layer in one of both encoders were used as the first states of each layer in the decoder.
The linear layer consists of 128 hidden nodes and 26 output softmax nodes.
In training, we applied dropout with the rate of 0.4 except on the inputs of the text encoder because of sparsity.
We used the Adam optimization algorithm \cite{kingma2015adam} with learning rate of 0.0001, $\beta_1$ = 0.9, $\beta_2$ = 0.999, and $\epsilon$ = $10^{-8}$.
The mini-batch size $N$ was set to 256 to maximize the use of memory capacity of two GTX 1080 Ti GPUs.
Models were trained for 150 epochs while evaluation for the development set was performed every epoch and finally the model having the highest development set AP was selected.
The baseline and proposed models were implemented in PyTorch \cite{paszke2017automatic}.

\begin{table}[t]
\caption{Coarse grid search results in terms of the development set AP.}
\begin{center}
\begin{tabular}{c|ccc}
\Xhline{3\arrayrulewidth}
\# of layers & 1 & 2 & \textbf{3} \\
 & 0.566 & 0.787 & \textbf{0.830} \\
\hline
\# of hidden units & 128 & 256 & \textbf{512} \\
 & 0.748 & 0.803 & \textbf{0.830} \\
\hline
$m$ & 0.3 & \textbf{0.5} & 0.7 \\
 & 0.802 & \textbf{0.830} & 0.800 \\
\hline
$N$ & 32 & 128 & \textbf{256} \\
 & 0.806 & 0.819 & \textbf{0.830} \\
\Xhline{3\arrayrulewidth}
\end{tabular}
\end{center}
\label{Tab:coarse}
\end{table}

\begin{figure}[t]
\begin{minipage}[b]{1.0\linewidth}
  \centering
  \centerline{\includegraphics[width=\linewidth]{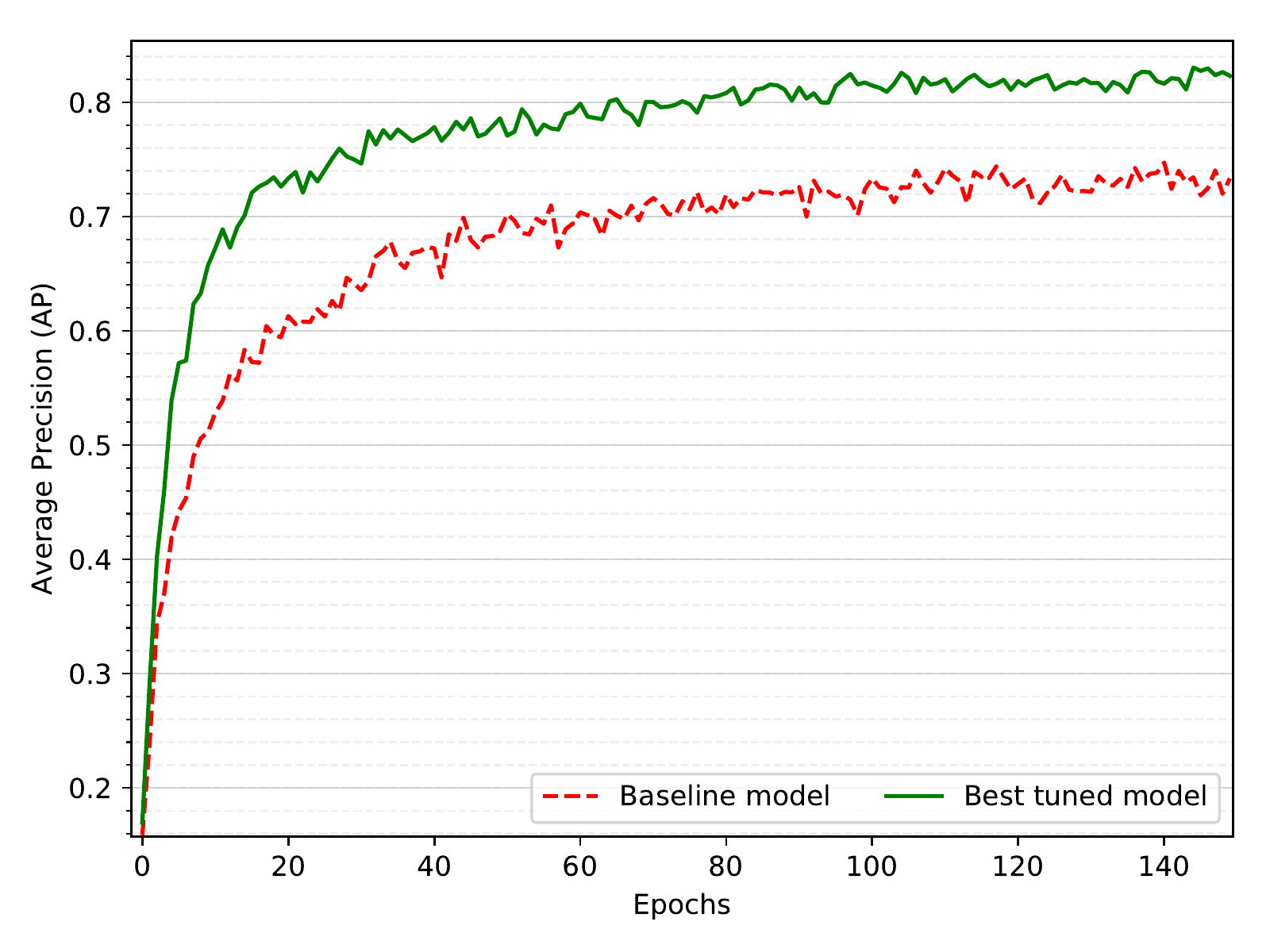}}
\end{minipage}
\caption{Progression of the development set AP for training baseline model and best tuned model on acoustic word discrimination.}
\label{fig:bestvsbaseline}
\end{figure}

\subsection{Model parameters tuning}
\label{ssec:subtuning}

To investigate the effectiveness of the additional shared decoder, we checked the development set AP for the initial model by changing the decoding loss weight $\alpha$ from 0, where the model is identical to the baseline, to 0.01, 0.05, 0.1, 0.3, and 0.6.
As can be seen in Table \ref{Tab:alphatable} and Fig.\ref{fig:alphavalue} (red dashed line), we found that the performance increases with $\alpha$ ranging from 0 to 0.1 and decreases with $\alpha >$ 0.1. 

Our proposed model was tuned by varying the parameters from the initial model of Sec.\ref{ssec:subsetup}.
We performed a coarse grid search in order of the number of LSTM layers over \{1, 2, \textbf{3}\}, the number of hidden units over \{128, 256, \textbf{512}\}, the margin $m$ over \{0.3, \textbf{0.5}, 0.7\}, the mini-batch size $N$ over \{32, 128, \textbf{256}\}, and the $\alpha$ over \{0.01, 0.05, \textbf{0.1}, 0.3, 0.6\}.
The numbers in bold are the selected parameters of the best tuned model.
The effect of varying the value of $\alpha$ is plotted in Fig.\ref{fig:alphavalue} (green line), which confirms the tendency of the model to perform best when $\alpha$ is 0.1.
The performance results of the coarse grid search are tabulated in Table \ref{Tab:coarse}.

Fig.\ref{fig:bestvsbaseline} shows the development set AP versus training epoch for the baseline model and best tuned model on acoustic word discrimination task.
We can observe that the performance gap of about 0.07 is maintained during the whole training process.

\begin{table}[t]
\caption{Final test set AP for the baseline multi-view approach and our proposed approach on acoustic word discrimination and cross-view word discrimination. }
\begin{center}
\begin{tabular}{c|c|cc}
\Xhline{3\arrayrulewidth}
\multicolumn{2}{c|}{\textbf{Model}} & \textbf{Acoustic} & \textbf{Cross-view} \\
\Xhline{3\arrayrulewidth}
\multicolumn{2}{c|}{Baseline \cite{he2016multi}} & 0.791 & 0.910 \\
\hline
\multirow{2}*{Proposed} & Initial & 0.841 & 0.935 \\
& Best tuned & \textbf{0.879} & \textbf{0.948} \\
\Xhline{3\arrayrulewidth}
\end{tabular}
\end{center}
\label{Tab:testresult}
\end{table}

\subsection{Results}
\label{ssec:subresults}

This paper does not include the results of DTW-based or single-view approaches.
Because in \cite{he2016multi}, the multi-view approach achieved significant performance improvement compared to previous approaches, we omit the unnecessary verification experiments.

In Table \ref{Tab:testresult}, we compare the final performance on the test set between the models of the baseline multi-view approach and our proposed approach.
We can clearly see that our proposed approach outperforms the baseline as in the previously observed results from the development set.
Our proposed approach achieved an AP of 0.879 on acoustic word discrimination task, which is 11.1\% relative improvement.
On cross-view word discrimination task, both approaches achieved a high AP over 0.9, because they were trained by adjusting the distance between cross-view embeddings with the multi-view triplet loss.
Here 4.2\% relative improvement was obtained, 0.948.

\subsection{Word-level speech recognition}
\label{ssec:subrecognition}

\begin{table}[t]
\caption{Examples of incorrectly decoded outputs and their original text labels for randomly selected words in the test set.}
\begin{center}
\begin{tabular}{c|c}
\Xhline{3\arrayrulewidth}
\textbf{Decoded output} & \makecell{remardin, hell, riiaa, digisting, tue, \\traik, ougust, blacforne, uv, genvrll, \\texaso, edecation, simene, turmantiu} \\
\hline
\textbf{Text label} & \makecell{remained, held, ryder, digesting, two, \\trade, august, blackburn, of, javelin, \\texans, education, symbol, terminate} \\
\Xhline{3\arrayrulewidth}
\end{tabular}
\end{center}
\label{Tab:decoded}
\end{table}

The decoder $h$ can be used in the test phase as well as in the training.
We extracted the acoustic embeddings of the test set and generated the decoded outputs from them to perform the word-level speech recognition.
Character error rate (CER) was calculated by comparing the decoded outputs with original text labels.
Of the words in the test set, 42.4\% CER was obtained for the in-vocabulary (IV) words seen in the training and 56.6\% CER was obtained for the OOV words.
Table \ref{Tab:decoded} shows incorrectly decoded ones among the examples for randomly selected words of the test set.
Although we obtained rather high CERs for OOV and also IV words, our proposed approach can be used as a pre-trained model for a word-level speech recognizer, judging from the plausible phonetic similarity which can be observed between the decoded outputs and actual text labels.

% --------------------------------------------------------------------------
%           5. Conclusion
% --------------------------------------------------------------------------
\section{CONCLUSION}
\label{sec:conclusion}

In this paper, we proposed an approach for jointly learning acoustic and text embeddings by introducing an advanced multi-view network architecture where an additional decoder is coupled and shared with the acoustic and text encoders.
In addition to multi-view triplet loss which allows embeddings to learn the phonetic similarity between words, decoding loss from encoder-decoder and autoencoder structures was also considered to maximize the representative capability and achieve normalization effect of embeddings.
We have found that consistent performance improvements can be obtained on word discrimination tasks.
Also, our proposed network architecture shows its potential to be used as a pre-trained model for a word-level speech recognizer.
Future work will consider a method that directly measures the phonetic similarity between text labels so that a large amount of text corpus can be used to train text embeddings in advance.

% --------------------------------------------------------------------------
%           6. Acknowledgement
% --------------------------------------------------------------------------
\section{ACKNOWLEDGEMENT}
\label{sec:acknowledgement}

This material is based upon work supported by the Ministry of Trade, Industry \& Energy (MOTIE, Korea) under Industrial Technology Innovation Program (No.10063424, Development of distant speech recognition and multi-task dialog processing technologies for in-door conversational robots).

\bibliographystyle{IEEEbib}
\bibliography{strings,refs}

\end{document}